# Score lists in multipartite hypertournaments

Shariefuddin Pirzada
Department of Mathematics, University of Kashmir, India, and
King Fahd University of Petroleum and Minerals, Saudi Arabia
email: sdpirzada@yahoo.co.in

Guofei Zhou
Department of Mathematics, Nanjing University, Nanjing, China
email: gfzhou@nju.edu.cn

Antal Iványi
Department of Computer Algebra
Eötvös Loránd University, Budapest, Hungary
email: tony@inf.elte.hu

**Abstract.** Given non-negative integers $n_i$ and $\alpha_i$ with $0 \le \alpha_i \le n_i$ ($i = 1, 2, \ldots, k$), an $[\alpha_1, \alpha_2, \ldots, \alpha_k]$-k-partite hypertournament on $\sum_1^k n_i$ vertices is a $(k+1)$-tuple $(U_1, U_2, \ldots, U_k, E)$, where $U_i$ are k vertex sets with $|U_i| = n_i$, and E is a set of $\sum_1^k \alpha_i$-tuples of vertices, called arcs, with exactly $\alpha_i$ vertices from $U_i$, such that any $\sum_1^k \alpha_i$ subset $\cup_1^k U_i'$ of $\cup_1^k U_i$, E contains exactly one of the $\left(\sum_1^k \alpha_i\right)! \sum_1^k \alpha_i$-tuples whose entries belong to $\cup_1^k U_i'$. We obtain necessary and sufficient conditions for k lists of non-negative integers in non-decreasing order to be the losing score lists and to be the score lists of some k-partite hypertournament.

## 1 Introduction

Hypergraphs are generalizations of graphs [1]. While edges of a graph are pairs of vertices of the graph, edges of a hypergraph are subsets of the vertex set, consisting of at least two vertices. An edge consisting of k vertices is called a k-edge. A k-hypergraph is a hypergraph all of whose edges are k-edges. A k-hypertournament is a complete k-hypergraph with each k-edge endowed with an orientation, that is, a linear arrangement of the vertices contained in the hyperedge. Instead of scores of vertices in a tournament, Zhou et al. [13] considered scores and losing scores of vertices in a k-hypertournament, and derived a result analogous to Landau's theorem [6]. The score $s(v_i)$ or $s_i$ of a vertex $v_i$ is the number of arcs containing $v_i$ and in which $v_i$ is not the last element, and the losing score $r(v_i)$ or $r_i$ of a vertex $v_i$ is the number of arcs containing $v_i$ and in which $v_i$ is the last element. The score sequence (losing score sequence) is formed by listing the scores (losing scores) in non-decreasing order.

The following characterizations of score sequences and losing score sequences in k-hypertournaments can be found in G. Zhou et al. [12].

**Theorem 1** *Given two positive integers* n *and* k *with* $n \geq k > 1$, *a non-decreasing sequence* $R = [r_1, r_2, \ldots, r_n]$ *of non-negative integers is a losing score sequence of some* k-*hypertournament if and only if for each* j,

$$\sum_{i=1}^{j} r_i \geq \binom{j}{k},$$

*with equality when* $j = n$.

**Theorem 2** *Given positive integers* n *and* k *with* $n \geq k > 1$, *a non-decreasing sequence* $S = [s_1, s_2, \ldots, s_n]$ *of non-negative integers is a score sequence of some* k-*hypertournament if and only if for each* j,

$$\sum_{i=1}^{j} s_i \geq j\binom{n-1}{k-1} + \binom{n-j}{k} - \binom{n}{k},$$

*with equality when* $j = n$.

Some recent work on the reconstruction of tournaments can be found in the papers due to A. Iványi [3, 4]. Some more results on k-hypertournaments can be found in [2, 5, 9, 10, 11, 13]. The analogous results of Theorem 1 and Theorem 2 for [h, k]-bipartite hypertournaments can be found in [7] and for $[\alpha, \beta, \gamma]$-tripartite hypertournaments in [8].

Throughout this paper i takes values from 1 to k and $j_i$ takes values from 1 to $n_i$, unless otherwise stated.

A k-partite hypergraph is a generalization of k-partite graph. Given non-negative integers $n_i$ and $\alpha_i$, ($i = 1, 2, \ldots, k$) with $n_i \geq \alpha_i \geq 0$ for each $i$, an $[\alpha_1, \alpha_2, \ldots, \alpha_k]$-k-partite hypertournament (or briefly k-partite hypertournament) $M$ of order $\sum_1^k n_i$ consists of $k$ vertex sets $U_i$ with $|U_i| = n_i$ for each $i$, ($1 \leq i \leq k$) together with an arc set $E$, a set of $\sum_1^k \alpha_i$-tuples of vertices, with exactly $\alpha_i$ vertices from $U_i$, called arcs such that any $\sum_1^k \alpha_i$ subset $\cup_1^k U_i'$ of $\cup_1^k U_i$, $E$ contains exactly one of the $\left(\sum_1^k \alpha_i\right) \sum_1^k \alpha_i$-tuples whose $\alpha_i$ entries belong to $U_i'$.

Let $e = (u_{11}, u_{12}, \ldots, u_{1\alpha_1}, u_{21}, u_{22}, \ldots, u_{2\alpha_2}, \ldots, u_{k1}, u_{k2}, \ldots, u_{k\alpha_k})$, with $u_{ij_i} \in U_i$ for each $i$, ($1 \leq i \leq k, 1 \leq j_i \leq \alpha_i$), be an arc in $M$ and let $h < t$, we let $e(u_{1h}, u_{1t})$ denote to be the new arc obtained from $e$ by interchanging $u_{1h}$ and $u_{1t}$ in $e$. An arc containing $\alpha_i$ vertices from $U_i$ for each $i$, ($1 \leq i \leq k$) is called an $(\alpha_1, \alpha_2, \ldots, \alpha_k)$-arc.

For a given vertex $u_{ij_i} \in U_i$ for each $i$, $1 \leq i \leq k$ and $1 \leq j_i \leq \alpha_i$, the score $d_M^+(u_{ij_i})$ (or simply $d^+(u_{ij_i})$) is the number of $\sum_1^k \alpha_i$-arcs containing $u_{ij_i}$ and in which $u_{ij_i}$ is not the last element. The losing score $d_M^-(u_{ij_i})$ (or simply $d^-(u_{ij_i})$) is the number of $\sum_1^k \alpha_i$-arcs containing $u_{ij_i}$ and in which $u_{ij_i}$ is the last element. By arranging the losing scores of each vertex set $U_i$ separately in non-decreasing order, we get $k$ lists called losing score lists of $M$ and these are denoted by $R_i = [r_{ij_i}]_{j_i=1}^{n_i}$ for each $i$, ($1 \leq i \leq k$). Similarly, by arranging the score lists of each vertex set $U_i$ separately in non-decreasing order, we get $k$ lists called score lists of $M$ which are denoted as $S_i = [s_{ij_i}]_{j_i=1}^{n_i}$ for each $i$ ($1 \leq i \leq k$).

## 2  Main results

The following two theorems are the main results.

**Theorem 3** *Given $k$ non-negative integers $n_i$ and $k$ non-negative integers $\alpha_i$ with $1 \leq \alpha_i \leq n_i$ for each $i$ ($1 \leq i \leq k$), the $k$ non-decreasing lists $R_i = [r_{ij_i}]_{j_i=1}^{n_i}$ of non-negative integers are the losing score lists of a $k$-partite hypertournament if and only if for each $p_i$ ($1 \leq i \leq k$) with $p_i \leq n_i$,*

$$\sum_{i=1}^{k} \sum_{j_i=1}^{p_i} r_{ij_i} \geq \prod_{i=1}^{k} \binom{p_i}{\alpha_i}, \tag{1}$$

*with equality when $p_i = n_i$ for each $i$ ($1 \leq i \leq k$).*

**Theorem 4** *Given $k$ non-negative integers $n_i$ and $k$ non-negative integers $\alpha_i$ with $0 \leq \alpha_i \leq n_i$ for each $i$ $(1 \leq i \leq k)$, the $k$ non-decreasing lists $S_i = [s_{ij_i}]_{j_i=1}^{n_i}$ of non-negative integers are the score lists of a $k$-partite hypertournament if and only if for each $p_i$, $(1 \leq i \leq k)$ with $p_i \leq n_i$*

$$\sum_{i=1}^{k}\sum_{j_i=1}^{p_i} s_{ij_i} \geq \left(\sum_{i=1}^{k} \frac{\alpha_i p_i}{n_i}\right)\left(\prod_{i=1}^{k}\binom{n_i}{\alpha_i}\right) + \prod_{i=1}^{k}\binom{n_i - p_i}{\alpha_i} - \prod_{i=1}^{k}\binom{n_i}{\alpha_i}, \quad (2)$$

*with equality when $p_i = n_i$ for each $i$ $(1 \leq i \leq k)$.*

We note that in a $k$-partite hypertournament $M$, there are exactly $\prod_{i=1}^{k}\binom{n_i}{\alpha_i}$ arcs and in each arc only one vertex is at the last entry. Therefore,

$$\sum_{i=1}^{k}\sum_{j_i=1}^{n_i} d_M^-(u_{ij_i}) = \prod_{i=1}^{k}\binom{n_i}{\alpha_i}.$$

In order to prove the above two theorems, we need the following Lemmas.

**Lemma 5** *If $M$ is a $k$-partite hypertournament of order $\sum_{1}^{k} n_i$ with score lists $S_i = [s_{ij_i}]_{j_i=1}^{n_i}$ for each $i$ $(1 \leq i \leq k)$, then*

$$\sum_{i=1}^{k}\sum_{j_i=1}^{n_i} s_{ij_i} = \left[\left(\sum_{1=1}^{k} \alpha_i\right) - 1\right]\prod_{i=1}^{k}\binom{n_i}{\alpha_i}.$$

**Proof.** We have $n_i \geq \alpha_i$ for each $i$ $(1 \leq i \leq k)$. If $r_{ij_i}$ is the losing score of $u_{ij_i} \in U_i$, then

$$\sum_{i=1}^{k}\sum_{j_i=1}^{n_i} r_{ij_i} = \prod_{i=1}^{k}\binom{n_i}{\alpha_i}.$$

The number of $[\alpha_i]_1^k$ arcs containing $u_{ij_i} \in U_i$ for each $i$, $(1 \leq i \leq k)$, and $1 \leq j_i \leq n_i$ is

$$\frac{\alpha_i}{n_i}\prod_{t=1}^{k}\binom{n_t}{\alpha_t}.$$

Thus,

$$\sum_{i=1}^{k}\sum_{j_i=1}^{n_i} s_{ij_i} = \sum_{i=1}^{k}\sum_{j_i=1}^{n_i} \left(\frac{\alpha_i}{n_i}\right)\prod_{1}^{k}\binom{n_t}{\alpha_t} - \binom{n_i}{\alpha_i}$$

$$= \left(\sum_{i=1}^{k}\alpha_i\right)\prod_{1}^{k}\binom{n_t}{\alpha_t} - \prod_{1}^{k}\binom{n_i}{\alpha_i}$$

$$= \left[\left(\sum_{1=1}^{k}\alpha_i\right) - 1\right]\prod_{1}^{k}\binom{n_i}{\alpha_i}.$$

□

**Lemma 6** *If* $R_i = [r_{ij_i}]_{j_i=1}^{n_i}$ $(1 \leq i \leq k)$ *are* $k$ *losing score lists of a* $k$-*partite hypertournament* $M$, *then there exists some* $h$ *with* $r_{1h} < \frac{\alpha_1}{n_1}\prod_{1}^{k}\binom{n_p}{\alpha_p}$ *so that* $R'_1 = [r_{11}, r_{12}, \ldots, r_{1h}+1, \ldots, r_{1n_1}]$, $R'_s = [r_{s1}, r_{s2}, \ldots, r_{st}-1, \ldots, r_{sn_s}]$ $(2 \leq s \leq k)$ *and* $R_i = [r_{ij_i}]_{j_i=1}^{n_i}$, $(2 \leq i \leq k)$, $i \neq s$ *are losing score lists of some* $k$-*partite hypertournament,* $t$ *is the largest integer such that* $r_{s(t-1)} < r_{st} = \ldots = r_{sn_s}$.

**Proof.** Let $R_i = [r_{ij_i}]_{j_i=1}^{n_i}$ $(1 \leq i \leq k)$ be losing score lists of a $k$-partite hypertournament $M$ with vertex sets $U_i = \{u_{i1}, u_{i2}, \ldots, u_{ij_i}\}$ so that $d^-(u_{ij_i}) = r_{ij_i}$ for each $i$ $(1 \leq i \leq k, 1 \leq j_i \leq n_i)$.

Let $h$ be the smallest integer such that

$$r_{11} = r_{12} = \ldots = r_{1h} < r_{1(h+1)} \leq \ldots \leq r_{1n_1}$$

and $t$ be the largest integer such that

$$r_{s1} \leq r_{s2} \leq \ldots \leq r_{s(t-1)} < r_{st} = \ldots = r_{sn_s}.$$

Now, let

$$R'_1 = [r_{11}, r_{12}, \ldots, r_{1h}+1, \ldots, r_{1n_1}],$$

$$R'_s = [r_{s1}, r_{s2}, \ldots, r_{st}-1, \ldots, r_{sn_s}]$$

$(2 \leq s \leq k)$, and $R_i = [r_{ij_i}]_{j_i=1}^{n_i}$, $(2 \leq i \leq k)$, $i \neq s$.

Clearly, $R'_1$ and $R'_s$ are both in non-decreasing order.

Since $r_{1h} < \frac{\alpha_1}{n_1}\prod_{1}^{k}\binom{n_p}{\alpha_p}$, there is at least one $[\alpha_i]_1^k$-arc $e$ containing both $u_{1h}$ and $u_{st}$ with $u_{st}$ as the last element in $e$, let $e' = (u_{1h}, u_{st})$. Clearly, $R'_1$, $R'_s$ and $R_i = [r_{ij_i}]_{j_i=1}^{n_i}$ for each $i$ $(2 \leq i \leq k)$, $i \neq s$ are the k losing score lists of $M' = (M - e) \cup e'$. □

The next observation follows from Lemma 6, and the proof can be easily established.

**Lemma 7** *Let $R_i = [r_{ij_i}]_{j_i=1}^{n_i}$, $(1 \leq i \leq k)$ be k non-decreasing sequences of non-negative integers satisfying (1). If $r_{1n_1} < \frac{\alpha_1}{n_1} \prod_1^k \binom{n_t}{\alpha_t}$, then there exists s and t $(2 \leq s \leq k)$, $1 \leq t \leq n_s$ such that $R'_1 = [r_{11}, r_{12}, \ldots, r_{1h} + 1, \ldots, r_{1n_1}]$, $R'_s = [r_{s1}, r_{s2}, \ldots, r_{st} - 1, \ldots, r_{sn_s}]$ and $R_i = [r_{ij_i}]_{j_i=1}^{n_i}$, $(2 \leq i \leq k)$, $i \neq s$ satisfy (1).*

**Proof of Theorem 3. Necessity.** Let $R_i$, $(1 \leq i \leq k)$ be the k losing score lists of a k-partite hypertournament $M(U_i, 1 \leq i \leq k)$. For any $p_i$ with $\alpha_i \leq p_i \leq n_i$, let $U'_i = \{u_{ij_i}\}_{j_i=1}^{p_i}(1 \leq i \leq k)$ be the sets of vertices such that $d^-(u_{ij_i}) = r_{ij_i}$ for each $1 \leq j_i \leq p_i$, $1 \leq i \leq k$. Let $M'$ be the k-partite hypertournament formed by $U'_i$ for each i $(1 \leq i \leq k)$.

Then,

$$\sum_{i=1}^{k}\sum_{j_i=1}^{p_i} r_{ij_i} \geq \sum_{i=1}^{k}\sum_{j_i=1}^{p_i} d^-_{M'}(u_{ij_i})$$

$$= \prod_1^k \binom{p_t}{\alpha_t}.$$

**Sufficiency.** We induct on $n_1$, keeping $n_2, \ldots, n_k$ fixed. For $n_1 = \alpha_1$, the result is obviously true. So, let $n_1 > \alpha_1$, and similarly $n_2 > \alpha_2, \ldots, n_k > \alpha_k$. Now,

$$r_{1n_1} = \sum_{i=1}^{k}\sum_{j_i=1}^{n_i} r_{ij_i} - \left(\sum_{j_1=1}^{n_1-1} r_{1j_1} + \sum_{i=2}^{k}\sum_{j_i=1}^{n_i} r_{ij_i}\right)$$

$$\leq \prod_1^k \binom{n_t}{\alpha_t} - \binom{n_1-1}{\alpha_1}\prod_2^k \binom{n_t}{\alpha_t}$$

$$= \left[\binom{n_1}{\alpha_1} - \binom{n_1-1}{\alpha_1}\right]\prod_2^k \binom{n_t}{\alpha_t}$$

$$= \binom{n_1-1}{\alpha_1-1}\prod_2^k \binom{n_t}{\alpha_t}.$$

We consider the following two cases.

**Case 1.** $r_{1n_1} = \binom{n_1-1}{\alpha_1-1} \prod_2^k \binom{n_t}{\alpha_t}$. Then,

$$\sum_{j_1=1}^{n_1-1} r_{1j_1} + \sum_{i=2}^{k} \sum_{j_i=1}^{n_i} r_{ij_i} = \sum_{i=1}^{k} \sum_{j_i=1}^{n_i} r_{ij_i} - r_{1n_1}$$

$$= \prod_1^k \binom{n_t}{\alpha_t} - \binom{n_1-1}{\alpha_1-1} \prod_2^k \binom{n_t}{\alpha_t}$$

$$= \left[\binom{n_1}{\alpha_1} - \binom{n_1-1}{\alpha_1-1}\right] \prod_2^k \binom{n_t}{\alpha_t}$$

$$= \binom{n_1-1}{\alpha_1} \prod_2^k \binom{n_t}{\alpha_t}.$$

By induction hypothesis $[r_{11}, r_{12}, \ldots, r_{1(n_1-1)}], R_2, \ldots, R_k$ are losing score lists of a k-partite hypertournament $M'(U_1', U_2, \ldots, U_k)$ of order $\left(\sum_{i=1}^k n_i\right) - 1$. Construct a k-partite hypertournament $M$ of order $\sum_{i=1}^k n_i$ as follows. In $M'$, let $U_1' = \{u_{11}, u_{12}, \ldots, u_{1(n_1-1)}\}, U_i = \{u_{ij_i}\}_{j_i=1}^{n_i}$ for each $i$, $(2 \leq i \leq k)$. Adding a new vertex $u_{1n_1}$ to $U_1'$, for each $\left(\sum_{i=1}^k \alpha_i\right)$-tuple containing $u_{1n_1}$, arrange $u_{1n_1}$ on the last entry. Denote $E_1$ to be the set of all these $\binom{n_1-1}{\alpha_1-1} \prod_2^k \binom{n_t}{\alpha_t}$ $\left(\sum_{i=1}^k \alpha_i\right)$-tuples. Let $E(M) = E(M') \cup E_1$. Clearly, $R_i$ for each $i$, $(1 \leq i \leq k)$ are the k losing score lists of $M$.

**Case 2.** $r_{1n_1} < \binom{n_1-1}{\alpha_1-1} \prod_2^k \binom{n_t}{\alpha_t}$.

Applying Lemma 7 repeatedly on $R_1$ and keeping each $R_i$, $(2 \leq i \leq k)$ fixed until we get a new non-decreasing list $R_1' = [r_{11}', r_{12}', \ldots, r_{1n_1}']$ in which now $r_{1n_1}' = \binom{n_1-1}{\alpha_1-1} \prod_2^k \binom{n_t}{\alpha_t}$. By Case 1, $R_1', R_i$ $(2 \leq i \leq k)$ are the losing score lists of a k-partite hypertournament. Now, apply Lemma 6 on $R_1'$, $R_i$ $(2 \leq i \leq k)$ repeatedly until we obtain the initial non-decreasing lists $R_i$ for each $i$ $(1 \leq i \leq k)$. Then by Lemma 6, $R_i$ for each $i$ $(1 \leq i \leq k)$ are the losing score lists of a k-partite hypertournament. □

**Proof of Theorem 4.** Let $S_i = [s_{ij_i}]_{j_i=1}^{n_i} (1 \leq i \leq k)$ be the k score lists of a k-partite hypertournament $M(U_i, 1 \leq i \leq k)$, where $U_i = \{u_{ij_i}\}_{j_i=1}^{n_i}$ with

$d^+_M(u_{ij_i}) = s_{ij_i}$, for each $i$, $(1 \leq i \leq k)$.   Clearly,
$d^+(u_{ij_i}) + d^-(u_{ij_i}) = \frac{\alpha_i}{n_i} \prod_1^k \binom{n_t}{\alpha_t}$, $(1 \leq i \leq k, 1 \leq j_i \leq n_i)$.

Let $r_{i(n_i+1-j_i)} = d^-(u_{ij_i})$, $(1 \leq i \leq k, 1 \leq j_i \leq n_i)$.

Then $R_i = [r_{ij_i}]_{j_i=1}^{n_i} (i = 1, 2, \ldots, k)$ are the $k$ losing score lists of $M$. Conversely, if $R_i$ for each $i$ $(1 \leq i \leq k)$ are the losing score lists of $M$, then $S_i$ for each $i$, $(1 \leq i \leq k)$ are the score lists of $M$. Thus, it is enough to show that conditions (1) and (2) are equivalent provided $s_{ij_i} + r_{i(n_i+1-j_i)} = \left(\frac{\alpha_i}{n_i}\right) \prod_1^k \binom{n_t}{\alpha_t}$, for each $i$ $(1 \leq i \leq k$ and $1 \leq j_i \leq n_i)$.

First assume (2) holds. Then,

$$\sum_{i=1}^k \sum_{j_i=1}^{p_i} r_{ij_i} = \sum_{i=1}^k \sum_{j_i=1}^{p_i} \left(\frac{\alpha_i}{n_i}\right) \left(\prod_1^k \binom{n_t}{\alpha_t}\right) - \sum_{i=1}^k \sum_{j_i=1}^{p_i} s_{i(n_i+1-j_i)}$$

$$= \sum_{i=1}^k \sum_{j_i=1}^{p_i} \left(\frac{\alpha_i}{n_i}\right) \left(\prod_1^k \binom{n_t}{\alpha_t}\right) - \left[\sum_{i=1}^k \sum_{j_i=1}^{n_i} r_{ij_i} - \sum_{i=1}^k \sum_{j_i=1}^{n_i-p_i} s_{ij_i}\right]$$

$$\geq \left[\sum_{i=1}^k \sum_{j_i=1}^{p_i} \left(\frac{\alpha_i}{n_i}\right) \left(\prod_1^k \binom{n_t}{\alpha_t}\right)\right]$$

$$- \left[\left(\left(\sum_1^k \alpha_i\right) - 1\right) \prod_1^k \binom{n_i}{\alpha_i}\right]$$

$$+ \sum_{i=1}^k (n_i - p_i) \left(\frac{\alpha_i}{n_i}\right) \prod_1^k \binom{n_t}{\alpha_t}$$

$$+ \prod_1^k \binom{n_i - (n_i - p_i)}{\alpha_i} - \prod_1^k \binom{n_i}{\alpha_i}$$

$$= \prod_1^k \binom{n_i}{\alpha_i},$$

with equality when $p_i = n_i$ for each $i$ $(1 \leq i \leq k)$.   Thus (1) holds.

Now, when (1) holds, using a similar argument as above, we can show that (2) holds. This completes the proof. $\square$

## Acknowledgements

The research of the third author was supported by the European Union and the European Social Fund under the grant agreement no. TÁMOP 4.2.1/B-09/1/KMR-2010-0003.